\documentclass[a4paper]{jpconf}

\newcommand{\DPV}{\Delta \partial V}
\newcommand{\Sp}{S_{\partial 1}}
\newcommand{\SpD}{S_{\partial 1 \ \DPV}}
\newcommand{\intc}[3]{\int_{#1} d^{#2}x \ #3}
\newcommand{\intct}[3]{\int_{#1} d^{#2}\tilde{x} \ #3}
\newcommand{\sqn}[1]{\sqrt{|#1|}}
\newcommand{\Lm}{\mathcal{L}}
\newcommand{\gah}[1]{\gamma^{#1}}
\newcommand{\gal}[1]{\gamma_{#1}}
\newcommand{\cfac}[1]{\frac{c^4}{#1 \pi G}}
\newcommand{\cface}[1]{\frac{c^4 \epsa{}}{#1 \pi G}}
\newcommand{\xpara}{x^{||}}
\newcommand{\tran}[2]{\frac{\partial #1}{\partial #2}}
\newcommand{\nline}{\nonumber \\ }
\newcommand{\bu}{\bar{u}}
\newcommand{\bv}{\bar{v}}
\newcommand{\hor}[2]{\mathcal{H}_{#1#2}}
\newcommand{\epsa}[1]{\varepsilon_A \ #1}
\newcommand{\inte}[2]{\epsa{\int_{#1} d^3x \ #2}}

\newcommand{\Jac}[2]{\big| \frac{e^2_{#1}}{e^2_{#2}}  \big|}
\newcommand{\sumk}{\sum_{k=1}^N}
\newcommand{\tT}[1]{\tilde{T}_{#1}}
\newcommand{\half}{\frac{1}{2}}
\newcommand{\Po}[1]{\mathcal{P}_{#1}}

\newcommand{\NdSEq}{N^c_{ab} \ d^3\Sigma_c = \epsa{N^{ab} \ d^3x}}

\begin{document}
\title{Multi-layer statistical gravity on the boundary}

\author{Pierre A Mandrin}

\address{Department of Physics, University of Zurich, \\Winterthurerstrasse 190, 8057 Z\"urich, Switzerland}

\ead{pierre.mandrin@uzh.ch}

\begin{abstract}
Starting from an important research path, we consider gravity as a collective phenomenon governed by statistical mechanics. While previous studies have focussed on the thermodynamic heat flow across a 2d-horizon as perceived by a single, accelerated observer, we evaluate here the number of microscopic states arising for multiple observers perceiving multiple horizons within foliations of the boundary of a space-time region. This yields a temperature-independent, Boltzmann-type ''entropy'' which is equivalent to the boundary action and which we call m-entropy. According to its statistical interpretation, the m-entropy distribution as a function of the gravitational field is maximum when Einstein's Field Equations hold. However, if the number of ''atoms of space'' is small, Einstein's Equations do not hold and no sharp geometry can be defined. On the other hand, the transition probability of microstates can be computed and can be interpreted as processes of a (alternative) model of quantum space-time.
\end{abstract}

%%%%%%%%% Introduction
\section{Introduction}
\label{intro}

The discovery of the thermodynamic properties of black holes by Bekenstein \cite{Bekenstein} and Hawking \cite{Hawking} has initiated an important research path, according to which the gravitational field equations acquire the same status as e.g. the equations of fluid mechanics, see former work by Jacobson \cite{Jacobson} and Padmanabhan \cite{Padmanabhan_2014}. The 2-surface on the null-horizon of the Rindler wedge is considered to be equivalent to a black hole horizon. Imposing the stationarity condition allows the definition of a local, observer-dependent temperature $T = \kappa / (2\pi)$ with surface gravity $\kappa$, as well as an entropy $S_{BHJP}=A/(4 L_p^2)$ with associated horizon surface area $A$, where $L_p$ is the Planck length.

%\paragraph*{}
In the present article, we consider an extension to a statistical model of the microscopic degrees of freedom. The statistical model is not restricted by (local) thermal equilibrium or by classical laws. Because the model is necessarily incompatible with a temperature concept, we have to introduce another type of ''entropy'' which we call the m-entropy $S$, it gathers missing information from multiple observers. The ''temperature''-independence makes $S$ very different from $S_{BHJP}$. As we will see, the underlying model can be interpreted as a quantum model of space-time on a dimensionally reduced space and can be used to compute transition probabilities of microscopic states.

%\paragraph*{}
Before presenting our model, we reformulate the action of general relativity so that it only acts on the boundary of a space-time region, and we then relate this action with the thermodynamic description of Rindler horizons. These preliminary steps are required in order to recognise later the statistical model capable of reproducing GR and, at the same time, identify a kind of alternative quantum model of space-time.

%%%%%%%%% The boundary action of general relativity (GR)
\section{The boundary action of general relativity (GR)}
\label{action}

Consider a compact space-time region $V$ with piece-wise smooth boundary $\partial V = \sum_A B_A$, where every component $B_A$ is non-null (time- or space-like). As we know, different expressions for the action equally describe GR. The Einstein-Hilbert action $S_H$ and, to first order, the ''covariant Einstein action'' $S_1$ are defined as (omitting the cosmological constant for simplicity) \cite{York}

\begin{eqnarray}
\label{S_H}
S_H & = & \frac{c^4}{16 \pi G} \int_V \sqrt{-g} \ R \ d^4 x + S_m, \\
\label{S_1}
S_1 & = & \frac{c^4}{16 \pi G} \bigg[ \int_V \sqrt{-g} \ R \ d^4 x - \sum_A 2 \varepsilon_A \int_{B_A} \sqrt{|\gamma|} \ K \ d^3x \bigg]  + S_m,
\end{eqnarray}

\noindent where $S_m$ is the matter action, $K = K_{ab} \gamma^{ab}$, $\gamma_{ab}$ is the induced metric on $\partial V$ with determinant $\gamma$, the exterior curvature on $\partial V$ with unit normal vector $n^a$ is
$K_{ab} = -\frac{1}{2} \Lm_\perp \gamma_{ab} = -g^{cd}\gamma_{da} \nabla_c n_b$,
$\Lm_\perp=\Lm_n$, and $\varepsilon_A = n^an_a$ on $B_A$. From now on, we ignore the matter term. Variation of (\ref{S_H}) and (\ref{S_1}) is known to yield a bulk term (leading to Einstein's field equations) and a boundary term

\begin{equation}
\label{delta_S_H_1}
\delta S_H\big|_{\partial V} =  \frac{c^4}{16 \pi G} \sum_A \varepsilon_A \int_{B_A} \gamma_{ab} \ \delta N^{ab} d^3x,
\quad \delta S_1\big|_{\partial V} = -\frac{c^4}{16 \pi G} \sum_A \varepsilon_A \int_{B_A} N^{ab} \ \delta\gamma_{ab} \ d^3x,
\end{equation}

\noindent respectively, where $N_{ab} = K_{ab} - K \gamma_{ab}$. We now combine

\begin{equation}
\label{S_d1}
S_{\partial 1} = \frac{1}{2} (S_H-S_1) = \frac{c^4}{16 \pi G} \sum_A \varepsilon_A \int_{B_A} \sqrt{|\gamma|} \ K \ d^3x, \qquad
S_{\Sigma 1} = \frac{1}{2} (S_H+S_1)
\end{equation}

\noindent with the variations ($S_{\partial 1}$ only acts on $\partial V$, and $\gamma_{ab}$ and $N^{ab}$ are now varied)

\begin{eqnarray}
\label{dS_d1}
\delta S_{\partial 1} & = & \frac{c^4}{32 \pi G} \sum_A \varepsilon_A \int_{B_A} \delta (N^{ab} \gamma_{ab}) \ d^3x , \\
\label{dS_S1}
\delta S_{\Sigma 1} & = & \frac{c^4}{32 \pi G} \sum_A \varepsilon_A \int_{B_A} \gamma_{ab} \ \delta N^{ab} - N^{ab} \ \delta\gamma_{ab} \ d^3x, \\
\label{saddles}
\delta S_{\partial 1}\big|_{\gamma_{ab}} & = & \delta S_{\Sigma 1}\big|_{g_{ab},\gamma_{ab}}, \quad
\delta S_{\partial 1}\big|_{N^{ab}} = -\delta S_{\Sigma 1}\big|_{g_{ab},N^{ab}},
\end{eqnarray}

\noindent where the suffix $\big|_Y$ means that $Y$ is fixed at its saddle point value of GR, and $S_{\partial 1}$ has its extrema at the same values of $\gamma_{ab}$ and $N^{ab}$ as $S_1$ and $S_H$, respectively. Moreover, by using

\begin{equation}
\label{holo}
2 \delta S_{\partial 1}\big|_{\gamma_{ab}} = \frac{c^4}{16 \pi G} \int_V \sqrt{-g} \ g_{ab} \  \delta R^{ab} \ d^4 x,
\end{equation}

\noindent the general bulk action ansatz $S_{\rm bulk} \sim \int_V \sqrt{-g} \ Q^{ab} \ g_{ab} \ d^4x$ with

\begin{equation}
\label{bulk}
\delta S_{\rm bulk} \sim \int_V \sqrt{-g} \ (Q^{ab} -\frac{1}{2}Q^c_c \ g^{ab}) \ \delta g_{ab} \ d^4x + \int_V \sqrt{-g} \ g_{ab} \ \delta Q^{ab} \ d^4 x
\end{equation}

\noindent and the condition $\delta S_{\rm bulk}\big|_{g_{ab}} = 2 \delta S_{\partial 1}\big|_{\gamma_{ab}} $ forces $R^{ab} \sim Q^{ab}$, so that the theory defined by the ''boundary action'' $S_{\partial 1}$ is equivalent to GR.

%%%%%%%%% Interpretation of the boundary action using multiple horizon thermodynamics
\section{Interpretation of the boundary action using multiple horizon thermodynamics}
\label{thermodynamics}

There have been arguments suggesting that the microscopic degrees of freedom of gravity live on 2-surfaces and there is a third dimension along which the microscopic state of the 2-surfaces is allowed to change. A well known reasoning at the root of the holographic principle is given by 't Hooft \cite{Dim_Red} and is based on general principles of quantum systems and on the properties of black holes. On the other hand, space-time may be described thermodynamically \cite{Jacobson, Padmanabhan_2014}. An accelerated observer associates a temperature $T$ and an entropy $S_{BHJP}$ to the perceived Rindler horizon. $T$ and $S_{BHJP}$ are well-defined at the limit of stationarity (or local equilibrium) of the horizon. As we see next, the product $TS_{BHJP}$ evaluated on a sequence of narrow null-horizon strips yields yet a different, Boltzmann-type entropy $S$ (we call it m-entropy), and $S \sim S_{\partial 1}$.

%\paragraph*{}
Consider a compact space-time region $V$ with non-null boundary $\partial V$. The boundary action $S_{\partial 1}$ (\ref{S_d1}) is equivalent to the boundary term in the notation of \cite{Padmanabhan_2014},

\begin{equation}
\label{S_d1_P}
S_{\partial 1} = \frac{c^4}{16 \pi G} \int_{\partial V} N^c_{ab} \ f^{ab} \ d^3\Sigma_c, \qquad f^{ab} = \sqrt{-g} \ g^{ab}, \qquad \NdSEq,
\end{equation}

\noindent $N^c_{ab}$ is the momentum conjugate to $f^{ab}$ and $d^3\Sigma_c$ is the 3-surface element covector. 

\noindent We now cut the boundary $\partial V$ into tiny cuboids $\DPV \subset B_A \subset \partial V$ and choose the length $L_e$ of each edge $e$ of $\DPV$ to satisfy $L_p \ll L_e <$ variation scale of $\gamma_{ab}$ and $K_{ab}$ on $\partial V$. We rewrite

\begin{equation}
\label{S_d1_sum}
S_{\partial 1} = \sum_{\DPV \subset \partial V} \SpD
\end{equation}

\noindent with

\begin{equation}
\label{S_d1_Delta_dV}
\SpD  =  \cfac{16} \int_{\DPV} N^c_{ab} \ f^{ab} \ d^3\Sigma_c
 =  \cfac{32} \ \inte{\DPV}{\sqn{\gamma} \ \gah{ab} \ \Lm_\perp \gal{ab}}.
\end{equation}

\noindent Without loss of generality, we restrict ourselves to the gauge $g_{0a}=g_{00}\delta^0_a$. Because $\gal{ab}$ is symmetric, it can be diagonalised using an orthogonal transformation, $x^a \rightarrow u^a_b x^b$, so that $\gal{ab} \rightarrow u_a^c \gal{cd} {(u^T)}^d_b$. The transformation $u^a_b$ is held fixed within $\DPV$ as $g_{ab}$ does not change significantly. If $e_{(a)}^I$ are the triads induced on $\partial V$, we write $\gal{ab} = e_{(a)}^I \eta_{IJ} e_{(b)}^J$, and $e_{(a)}^I$ is diagonal. The transformation $u^a_b$ has unit Jacobian determinant and (\ref{S_d1_Delta_dV}) transforms to

\begin{equation}
\label{diag}
\SpD = \cfac{16} \ \inte{\DPV}{\sqn{\gamma} \ \sigma}, \qquad
\sigma = \frac{1}{2} u^a_c \gah{cd} {(u^T)}^d_b \ \Lm_\perp u_b^e \gal{ef} {(u^T)}^f_a = \frac{1}{2} \gah{ab} \ \Lm_\perp \gal{ab}.
\end{equation}

\noindent Therefore, $\gal{ab}$ can be assumed to be diagonal without loss of generality.  We reexpress (\ref{diag}) in terms of null coordinates $(\tilde{x}^a)=(\tilde{x}^1=u,\tilde{x}^2=v, x^3,x^4)$ instead of $(x^a)=(x^\perp,\xpara, x^3,x^4)$, where $\xpara$ is tangent to $\partial V$ and $x^3,x^4$ are both spatial components. The range of integration in (\ref{diag}) approximately splits into an interval $\mathcal{I}=[\xpara_-,\xpara_+]$ and a piece of 2-surface $A(\xpara)$:

\begin{equation}
\label{range}
\intc{\DPV}{3}{} = \intc{A(\xpara)}{2}{} \int_\mathcal{I} d\xpara.
\end{equation}

\noindent If we assume that the classical spatial resolution of $\xpara$ is no better than $\Delta \xpara$, (\ref{diag}) can be written in terms of a sum of $N$ strips of width $\Delta \xpara$ with $N = (\xpara_+-\xpara_-) / \Delta \xpara$:

\begin{equation}
\label{S_d1DdV_foliated}
S_{\partial 1 \ \DPV} = \cface{16} \int_{\mathcal{I}} d\xpara \ \intc{A(\xpara_k)}{2}{\sqn{\gamma} \ \sigma}
= \cface{16} \sumk \intc{A(\xpara_k)}{2}{\sqn{\gamma} \ \Delta \xpara \ \sigma},
\end{equation}

\noindent where $\xpara_k=\xpara_-+k\Delta \xpara$ and $A(\xpara_k)$ is the intersection of $\DPV$ with the ($x^3,x^4$)-surface at $\xpara=\xpara_k$. Because the triads are diagonal, $e^I_a$ also are diagonal, and we can rewrite (with $A=3,4$)

\begin{eqnarray}
\sigma & = & e^{(a)}_I \eta^{IJ} e^{(b)}_J e^K_{(b)} \eta_{KL} \Lm_\perp e^L_{(a)} = e^{(a)}_I \eta^{IJ} e^b_J e^K_b \eta_{KL} \Lm_\perp e^L_{(a)} = e^{(a)}_I  \Lm_\perp e^I_{(a)} = e^A_I  \Lm_\perp e^I_A + e^{||}_I  \Lm_\perp e^I_{||} \nline
& = & e^A_I  (\tran{v}{x^\perp} \ \Lm_v e^I_A + \tran{u}{x^\perp}  \ \Lm_u e^I_A) + (e^u_I \ \Lm_v + e^v_I \ \Lm_u) (e^I_u \tran{u}{\xpara} + e^I_v \tran{v}{\xpara}) \nline
\label{sigma_triad} 
& = & e^A_I  (\tran{u}{\xpara} \ \Lm_v e^I_A + \tran{v}{\xpara}  \ \Lm_u e^I_A) \ +
e^u_I \ \Lm_v e^I_u \tran{u}{\xpara} + e^v_I \ \Lm_u e^I_v \tran{v}{\xpara}
+ e^u_I \ \Lm_{||} e^I_v + e^v_I \ \Lm_{||} e^I_u. 
\end{eqnarray}

\noindent The last two terms of (\ref{sigma_triad}) sum up to zero because $e^v_I \ \Lm_{||} e^I_u = -e^I_u \ \Lm_{||} e^v_I = e^u_I \ \Lm_{||} e^I_v$ (by repeatedly using Leibnitz' rule and using $e^v_I e^I_u = \delta^v_u$, $e^v_I e^J_v = \delta^J_I$). With $\Delta \xpara = \Delta u \tran{\xpara}{u} = \Delta v \tran{\xpara}{v}$, we obtain

\begin{eqnarray}
\sigma \ \Delta \xpara & = & \Delta u \ (e^A_I \ \Lm_v e^I_A + e^u_I \ \Lm_v e^I_u) + \Delta v \ (e^A_I \ \Lm_u e^I_A + e^v_I \ \Lm_u e^I_v) \nline
\label{eq:sigma_dxpara}
& = & \Delta u \ e^{(a)_u}_I \ \Lm_v e^I_{(a)_u} + \Delta v \ e^{(a)_v}_I \ \Lm_u e^I_{(a)_v},
\end{eqnarray}

\noindent where $e^I_{(a)_w}$ are the triads induced on the hypersurface $w =$ constant with $w=u,v$. We diagonalise $e^I_{(a)_w}$ again with unit Jacobian and expand back to metric notation, to obtain

\begin{eqnarray}
& & \SpD = \cface{32} \sumk \intct{A(\xpara_k)}{2}{\bigg[ \sqn{\gal{(u)}}\Jac{||}{u} \ \Delta u \ \gah{ab}_{(u)} \ \Lm_v \gal{ab}^{(u)} + \sqn{\gal{(v)}}\Jac{||}{v} \ \Delta v \ \gah{ab}_{(v)} \ \Lm_u \gal{ab}^{(v)} \bigg]} \nline
& & = \cface{32} \sumk \bigg[ \int_{\Delta u} \intct{A(\xpara_k)}{3}{\sqn{\gal{(u)}}\Jac{||}{u} \ \gah{ab}_{(u)} \ \Lm_v \gal{ab}^{(u)}} 
\nline
\label{general_null}
& & + \int_{\Delta v} \intct{A(\xpara_k)}{3}{\sqn{\gal{(v)}}\Jac{||}{v} \ \gah{ab}_{(v)} \ \Lm_u \gal{ab}^{(v)}} \bigg],
\end{eqnarray}

\noindent where $\gal{ab}^{(w)}$ and $\gah{ab}_{(w)}$ are the metric and inverse metric induced on the hypersurface $w =$ constant, respectively. To link this result with horizon thermodynamics, we need to introduce locally flat coordinates $(\bar{t},\bar{x},\ldots)$ as well. The flat null coordinates $\bv$, $\bu$ are affine null parameters on the null surfaces $\bu=$ constant and $\bv=$ constant, respectively, which can locally be approximated by the null surfaces defined by the null strips. Then, for any general null-surface $\Sigma$, we have the identity \cite{Majhi_Padmanabhan}

\begin{equation}
\label{S_d1_MP}
\frac{c^4}{16 \pi G} \int_{\Sigma} N^c_{ab} \ f^{ab} \ d^3\Sigma_c = \int_{\Sigma} T \ s \ d\lambda \ d^2x,
\end{equation}

\noindent where $s$ is the surface density of the entropy $S_{BHJP}$ and $T$ the temperature of the horizon parametrised by $\lambda = \bu$ or $\bv$, as seen by the corresponding Rindler observer. We consider two groups of equally accelerated observers, one observer per strip, one group for the strips $\bu=$ constant observes the temperature $T_u$ and one group for the strips $\bv=$ constant observes the temperature $T_v$, respectively. Then, (\ref{general_null}) can also be written as

\begin{eqnarray}
\SpD & = & \half \sumk \bigg[ \int_{\Sigma_{(u)}(\xpara_k)} \tT{u} \ s \ du \ d^2x + \int_{\Sigma_{(v)}(\xpara_k)} \tT{v} \ s \ dv \ d^2x \bigg] \nline
\label{multi_observers}
& = & \half \sumk \int_{A(\xpara_k)} \big[ T_u \ s \ \Delta u  + T_v \ s \ \Delta v \big] \ d^2x,
\end{eqnarray}

\noindent where $\tT{w} = \Jac{||}{w} \ T_w$ and $w = u,v$.

%\paragraph*{} 
What is the interpretation of (\ref{multi_observers}) from the point of view of Rindler observers? To observe a temperature requires enough proper observer time to capture Unruh radiation and analyse its spectrum, i.e. the process lasts between two well-distinguishable times. Alternatively, we could also argue for spatially distinguishable points.

We can easily analyse the effect of having two neighbouring, identically accellerated Rindler observers $A$ and $B$ separated by the vector $(-\Delta \bu, \Delta \bv, 0, 0)$. The Rindler wedges of $A$ and $B$ are given by $\bar{x}_A > |\bar{t}_A|$ and $\bar{x}_B > |\bar{t}_B|$, respectively. We then define the past horizons $\hor{A}{-}$ and $\hor{B}{-}$ ($\bv=$ constant), the future horizons $\hor{A}{+}$ and $\hor{B}{+}$ ($\bu=$ constant), and the 2-surfaces $\Po{A}: \bar{t}_A=\bar{x}_A=0$ and $\Po{B}: \bar{t}_B=\bar{x}_B=0$. If the distance $\Delta \bar{x}^{||}$ between $\Po{A}$ and $\Po{B}$ is time-like and $A$ precedes $B$, $\hor{A}{+}$ and $\hor{B}{-}$ intersect precisely on a 2-surface $\Po{A'}$ at an affine parameter distance $\Delta \bv$ from $\Po{A}$ and at $\Delta \bu$ from $\Po{B}$. 

If $\xpara$ is space-like, we choose $B$ and its (positive) wedge so that it lies within the wedge of $A$ and then repeat the above argument.

In any case, the extended physical system made of both observers $A$ and $B$ must see twice as much horizon area (along $\hor{A}{-}$ and $\hor{B}{-}$) for values $\bu < \Delta \bu$ as would a single observer $A$ along $\hor{A}{-}$. For this reason, we need to sum up the entropies for all the null-strips contained in (\ref{multi_observers}) in order to describe the thermodynamics of the full multiple-observer system of the region $\DPV$.

%\paragraph*{} 
We shall now have a closer look at the (observer dependent) parameters $T_u, T_v$. Depending on these values, the observers will have very different records: 

\begin{enumerate}
\item If $\Delta \bar{v} > (\sqrt{2}T_v)^{-1}$, $\hor{B}{-}$ masks the preceding observer $A$ during its motion in the past to $\Po{A}$. Therefore, $B$ has no chance to notice the presence of $\hor{A}{}$. On the other hand, we can squeeze the sequence of wedges along $\xpara$ up to $\Delta \bar{v} = (\sqrt{2}T_v)^{-1}$ while keeping $A$ masked in the past to $\Po{A}$. This will increase the number of wedges and observers accordingly. However, the spacings below the initial $\Delta \xpara$ are to small for being resolved, by assumption, and not all horizon surfaces can be distinguished from each other. For wedges with space-like spacings $\Delta \xpara$, the logic of masking preceding observers and of resolution works in the same manner.
\item If $\Delta \bar{v} < (\sqrt{2}T_v)^{-1}$, the masking of observers in every wedge fails. In order to recover it, we have to stretch up the sequence of wedges to $\Delta \bar{v} = (\sqrt{2}T_v)^{-1}$. While stretching the sequence, the spacings of the wedges remain well resolved.
\end{enumerate}

\noindent The conditions

\begin{equation}
\label{temp_condition}
 \Delta \bu = (\sqrt{2}T_u)^{-1} \quad \Delta \bv = (\sqrt{2}T_v)^{-1}
\end{equation}

\noindent are therefore relevant when we ask how much entropy is made inaccessible to any of the observers. In fact, imposing (\ref{temp_condition}) dramatically simplifies (\ref{multi_observers}) to the following identity:

\begin{equation}
\label{special_case}
\Sp = 2^{-1/2}  \sumk  {S_{BHJP}}\big|_{\xpara_k; \ \bu < \Delta \bu = (\sqrt{2}T_u)^{-1}; \ \bv < \Delta \bv =  (\sqrt{2}T_v)^{-1}}.
\end{equation}

\noindent (\ref{special_case}) tells us that $\Sp$ reduces to a sum of stacked entropies $S_{BHJP}$ under stationary condition and for observers simultaneously resolving the loci of the stacked surfaces and maximising the amount of inaccessible information. For normal space, $\Sp$ provides an extension of $S_{BHJP}$ to multiple-observer systems as given by the usual boundary of an arbitrary compact space-time region $V$. Note that $\Sp$ also differs from $S_{BHJP}$ in that its value does not depend on the choice of a specific horizon temperature. In order to use an unambiguous terminology, we shall call $\Sp$ the ''m-entropy'' (for multiple observer entropy), in contrast to the single-observer entropy.

%%%%%%%%% Statistical interpretation of the boundary action

\section{Statistical interpretation of the boundary action}
\label{sec:statistics}

Now that we have found an interpretation of $\Sp$ as a generalised form of entropy of the space-time geometry, we may look for a general statistical description of it. According to the above findings, the region $\DPV$ can be foliated into a certain maximum number $N$ of pieces of space-like 2d-surfaces of aria $A_k$ restricted on $\DPV$ ($k=1, \ldots, N$). The number $N$ is given by the best resolution we can achieve in a classical frame-work. The 2d-surfaces satisfy $x^b =$ constant ($b=2,3$ or 4).
Each 2d-surface corresponds either to $\Po{A}$ or to $\Po{A'}$ (where $A$ and $A'$ are observers). Because the difference $\Delta x^\perp$ between $\Po{A}$ and the nearest $\Po{A'}$ can hardly be resolved and is cancelled in the step to the following $\Po{B}$, we can neglect $\Delta x^\perp$ (simpler structure). Moreover, $\DPV$ is a cuboid, all $A_k$ have the same size and we shorten $A_k \rightarrow A$.

%\paragraph*{}
We shall first consider the case for which $\DPV$ is space-like. Then, there is no preferred orientation $b$ across which to foliate. Rather, we could choose any spatial direction in 3d-space. We are also free to choose a representation with Minkowski index $K=2,3,4$ so that one value of $K$ represents the orientation of the foliation. It is convienient to write $A_K \sim n_K$ (note that $K$ is not used as a tensorial index).

%\paragraph*{}
Due to Bekenstein's interpretation of a horizon in terms of entropy, $A_K$ represents $n_K$ space-time atoms which can each have $n_q$ different states $|q\rangle$ ($n_q$ is a fixed positive integer number), yielding a total of 

\begin{equation}
\label{area}
p_K = n_q^{n_K}
\end{equation}

\noindent possible states per layer. All states $|q\rangle$ have the same probability to occur, and the number $p_K$ of states is related to a Boltzmann-type entropy $\sigma_K$ per layer of area $A_K$ via

\begin{equation}
\label{A_K}
\sigma_K = \ln{p_K}, \quad A_K = 4 L_p^2 \ln{p_K}.
\end{equation}

%\paragraph*{}
If we foliate $\DPV$ across orientation $K$ into $N_K$ layers, the number of microscopic states is

\begin{equation}
\label{eq:Omega}
\Omega_K(\DPV) = p_K^{N_K} = \exp{s_K},
\end{equation}

\noindent and
\begin{equation}
\label{eq:NK_nK}
s_K = N_K \sigma_K = N_K \ln{p_K} = N_K n_K \ln{n_q}
\end{equation}

\noindent is the Boltzmann-type entropy per volume $\DPV$ (or ''entropy density''). As previously mentioned, the distance $\Delta x^K$ between two consecutive layers is the smallest accessible spatial resolution. However, one expects $\Delta x^K$ to have a fundamental lower bound of the order of the Planck length $L_P$. $N_K$ can be estimated using the relation $N_K \approx k_K l^K$, where $l^K$ is the (Minkowski coordinate) size of $\DPV$ along $x^K$ ($l^J$ is a vector with Minkowski index) and the ''narrowness parameter'' $k_K$ is locally constant and of order $\le L_p^{-1}$. 

%\paragraph*{}
We can rewrite $s_K$ to a convenient form for direct comparison with the integrand of (\ref{S_d1_Delta_dV}):

\begin{equation}
\label{eq:sdens_triads}
s_K = k_K l^K \ln{p_K} = k_K l^j e^K_j \ln{p_K} = \sqrt{\gamma} \kappa^j_J e^J_j,
\end{equation}

\noindent with the definition $\sqrt{\gamma} \kappa^j_J = k_K l^j \ln{p_K} \delta_J^K$ and reintroducing the Einstein summation convention for the repeated index $J$ in the last expression of (\ref{eq:sdens_triads}). Finally, the total m-entropy coincides with $\Sp$ in (\ref{S_d1}), as far as the space-like parts $A$ of the boundary $\partial V$ are concerned:

\begin{eqnarray}
S_A & = & \sum_{\DPV \in A} \sqrt{\gamma} \kappa^j_K e^K_j,  \nline
\sqrt{\gamma} \kappa^j_K & = &  \cfac{16} \ \inte{\DPV}{K^j_K},  \nline
\label{eq:S_triads}
K^j_K & = & K^{ji} e_i^I \eta_{IK}.
\end{eqnarray}

\noindent From (\ref{eq:S_triads}), we obtain the value $k_K$ for every orientation $K$. More importantly, (\ref{eq:S_triads}) relates $K^j_K$ to the microscopic degrees of freedom of space.

%\paragraph*{}
If $\DPV$ is time-like, the foliation method straight-forwardly applies to the time-like orientation ($K=2$). But does $K^j_K$ represent the microscopic degrees of freedom as well? To reproduce $\Sp$ (\ref{S_d1}) for time-like parts of $\partial V$ as well, we would need to allow all 3 orientations, $K=2,3,4$, for the computation of the m-entropy, so that time-like layers would represent space-time atoms as well. In fact, we can support even this hypothesis. While crossing a Rindler horizon, Rindler time becomes a spatial Rindler coordinate, and vice-versa. We arrive in a time-like wedge and transform to ''new'' Minkowski-coordinates. We then choose a different spatial direction for the acceleration of an observer, then cross the horizon again, and finally recover Minkowski coordinates with a different time direction than before, but fully equivalent to the initial Minkowski coordinates. We can, however, recover the the initial horizon area which is now time-like, while the Rindler horizons can be recovered via a substitution of the form $\bar{x}^1 \rightarrow {\rm i}\bar{x}^1$. With this concept, time-like volumes $\DPV$ can be foliated across all 3 orientations as well.

%\paragraph*{}
To summarise, the boundary action $\Sp$ is proportional to the total m-entropy $S$ of the gravitational degrees of freedom on $\partial V$. $\Sp$ can be interpreted as a multiple entropy of layers of microscopic degrees of freedom of space-time.

%\paragraph*{}
By virtue of the above construction, the microstates of gravity are given by arbitrary states $|q\rangle$ on every layer in any orientation, and they are not restricted to states satisfying the field equations of GR. Nevertheless, in order to obtain the macroscopic state with the highest probability, we maximise the distribution of m-entropy $S$ as a function of $e^I_a$, i.e. we set the variation of $S$ to zero with respect to $e^I_a$. This is equivalent to applying Hamilton's variation principle to the Einstein-Hilbert action and yields Einstein's Equations. 
Some kind of analogy between the gravitational action and the notion of ''entropy'' has been suggested earlier \cite{Mandrin1, Mandrin2, Munkhammar}. For quantum mechanics, such a relationship also has been conjectured earlier \cite{Lisi}. 

%\paragraph*{}
The procedure described until now leads to a vacuum space-time geometry. However, we can generate a non-vacuum geometry by constraining the space of admissible microstates. We shortly sketch one method which consists in constraining the macroscopic quantities $e^I_a$, $\Lm_\perp e^I_a$ so that $\delta S \ne 0$ follows. We illustrate this procedure for two different sources of information about the space-time geometry.

\begin{enumerate}
\item We impose sources of gravitational field with unknown space-time distribution and specify how they determine the gravitational field. These are the holographic dual of the matter fields described by a Lagrangian $\Lm_m(a^j,f_j)$ with a set of free constants $a^j$ and fixed functions $f_j(e^\Delta_\mu,x^\mu)$. 
\item We impose part of the space-time distribution of the gravitational field (or matter fields) directly. This corresponds to the information we could obtain from local measurements.
\end{enumerate}

In both cases, the Lagrange multiplier method can be used so that we must add to $S$ an additional term, we call it $\mu(x^a)$:

\begin{equation}
\label{eq:S_t}
S_t = S + \intc{\delta V}{3}{\sqn{\gamma} \ \mu}
\end{equation}

\noindent If the expression $S_t$ is maximised instead of $S$ ($\delta S_t = 0$), it yields non-trivial classical solutions of GR under specific conditions and/or with non-vanishing matter.

%%%%%%%%% The alternative quantum treatment
\section{The alternative quantum treatment}
\label{sec:quantum_gravity}

The above statistical interpretation of the boundary action has lead us to evaluate a discrete number of possible microstates. We can consider two limits.

%\paragraph*{}
The classical limit is $N_K, p_K \rightarrow \infty$, and the classical dynamics are obtained by maximising $S_t(N_K(\DPV_m), p_K(\DPV_m), \mu(\DPV_m))$ as a function of the boundary gravitational field $e^I_a(l^a(\DPV_m); N_K(\DPV_m))), p_K(\DPV_m)))$ -- which we obtain from the relations (\ref{A_K}) -- and its Lie derivative $\Lm_\perp e^I_a$, where the index $m$ specifies every cuboid. In these terms, the classical limit is characterised by a sharp maximum of the m-entropy.

%\paragraph*{}
The quantum limit occurs for $N_K, p_K$ not much greater than 1. Although the counting of single states becomes relevant, we nevertheless consider the gravitational field $e^I_a$ again as a variable.
The distribution of $S$ does not have a sharp maximum as a function of the gravitational field. Neighbouring field values yield similar magnitudes. As the gravitational field is fully characterised by the quantum numbers $N_K, p_K$, we define the microscopic configuration $\mathcal{G}$ as the number set

\begin{equation}
\label{eq:param}
\mathcal{G} = (e^K_k(\DPV_m), N_K(\DPV_m), p_K(\DPV_m)).
\end{equation}

\noindent Given a configuration $\mathcal{G}$ for which some macroscopic quantity $\pi$ differs at some location $x^\mu$ by a small difference $D_\pi$ from its value at the maximum m-entropy configuration $\mathcal{G}_{\rm max}$, we have 

\begin{equation}
\label{eq:displace}
S_t(\mathcal{G})-S_t(\mathcal{G}_{\rm max}) \sim -D_\pi^2 + \mathcal{O}(D_\pi^3).
\end{equation} 

\noindent We can thus approximate the distribution $\Omega(\mathcal{G})$ around the maximum by a Gaussian distribution, and the standard deviation of $\pi$ determines the uncertainty of $\pi$. In this way, $e^K_k$ acquires an uncertainty, and this affects any space-time distance.
In this way, the statistical interpretation of gravity leads to a quantum uncertainty of the geometry.

%\paragraph*{}
Finally, we can analyse different microscopic states within a macroscopic state. Similarly to the $n$-point function computations of quantum field theory, we can evaluate the ''transition'' probability between $n$ local quantum geometries at the 3d-locations given by $\DPV_m$, $m=1\ldots n$, which defines a configuration $\mathcal{G}_m$. For example, let every number $p_K^m$ of microscopic states per layer of $\mathcal{G}_m$ at $\DPV_m$ be restricted by the quantum channel 

\begin{equation}
\label{eq:pkm}
p_K^{m \ ({\rm mod}\ n) \ + 1} = p_K^m + \Delta p_K^{m \ ({\rm mod}\ n) \ + 1}
\end{equation}

\noindent with $\Delta p_K^m$ being integers not far from 0 (given for this channel) and $\sum_m \Delta p_K^m = 0$. Then, taking into account all possible quantum configurations $\mathcal{G}_l$ within the narrow range of macroscopically undistinguishable $e^K_k$-values, one of which is $\mathcal{G}_m$, we can compute the transition probability as

\begin{equation}
\label{eq:n_point} 
p(\DPV_m, \Delta p_K^m) = \frac{\Omega(\mathcal{G}_m)}{\sum_l \Omega(\mathcal{G}_l)}, \qquad \Omega(\mathcal{G}) = exp S_t \big|_\mathcal{G}.
\end{equation}
  
%\paragraph*{}
Expressions of the form (\ref{eq:n_point}) are the starting point for predictions on outcome rates of quantum processes. Notice that the denominator of (\ref{eq:n_point}) cannot be interpreted as a ''partition function'' because no relaxation can take place in 3d-quantum space.

%%%%%%%%%%%%% Conclusions
\section{\label{sec:conclusion}Conclusions}

By considering a statistical interpretation of gravity on the boundary of a space-time region, a connection between the classical and a quantum formulation of gravity has been identified. The statistical interpretation has been obtained from a multiple observer perspective of Rindler horizon thermodynamics. Within the statistical frame-work, the microscopic degrees of freedom can have arbitrary states, but not every macroscopic state (or geometry) is likely to occur. The geometries with highest probability satisfy the classical theory (GR). In the quantum regime (small numbers of degrees of freedom), the geometric parameters are not sharp. The transition probabilities of microscopic states can be computed as a weighted sum over microscopic configurations.

%%%%%%%%%%%%%%%% Acknowledgements
\ack
I would like to thank Gino Isidori for fruitful discussions and Philippe Jetzer for hospitality at University of Zurich.

%%%%%%%%%%%%%%%%%%%
%%%%%%%%%%%%%%%%%%%

\end{document}